\documentclass[12pt]{iopart}

\usepackage{graphicx}
\usepackage{dcolumn}
\usepackage{bm}
\usepackage{mathbbol}
\usepackage{times}
\usepackage{graphicx}
\usepackage{bm}
\usepackage{color}
\usepackage{lipsum}
\usepackage{mcite}

\def\be{\begin{equation}}
\def\ee{\end{equation}}
\def\ba{\begin{eqnarray}}
\def\ea{\end{eqnarray}}

\newcommand{\gsim}{\raisebox{-0.13cm}{~\shortstack{$>$ \\[-0.07cm] $\sim$}}  ~}

\begin{document}
\title{From few- to many-body quantum systems}

\author{Mauro Schiulaz, Marco T\'avora, and Lea F. Santos}
\address{Department of Physics, Yeshiva University, New York, New York 10016, USA}

\date{\today}

\begin{abstract}
How many particles are necessary to make a quantum system many-body? To answer this question, we take as reference for the many-body limit a quantum system at half-filling and compare its properties with those of a system with $N$ particles, gradually increasing $N$ from 1. We show that convergence for the static properties of the system with few particles to the many-body limit is fast. For $N \gsim 4$, the density of states is already very close to Gaussian and signatures of many-body quantum chaos, such as level repulsion and fully extended eigenstates, become evident. The dynamics, on the other hand, depend on the initial state and time scale. In dilute systems, as the particles move away from each other, the entropy growth changes in time from linear, as typical for many-body systems, to logarithmic.
\end{abstract}


\maketitle

\section{Introduction}
Sorites paradoxes are usually illustrated by our inability to precise how many grains of sand constitutes a heap. One grain is not enough, nor two or three grains. But there is a point, even though not well marked, above which the collection of grains can be called a heap. The same question may be extended to the quantum limit: how many particles compose a many-body quantum system? Despite being a natural question, especially given the widespread theoretical and experimental interest in many-body quantum systems, the available related literature is surprisingly limited. While topics such as the onset of thermalization, the metal-insulator transition, and the scrambling of quantum information in interacting many-body quantum systems permeate studies in condensed matter physics, atomic and molecular physics, high energy physics, and quantum information theory, very little attention has been devoted to determining the minimum number of particles necessary to perform such studies.

Experiments with cold atoms, ion traps, and photon-based platforms are promising testbeds for addressing this point. In these experiments, the number of particles can be adjusted as desired~\cite{Zinner2016EPJWebConference}, which allows for studying how many-body effects emerge as the number of particles increases~\cite{Serwane2011,Wenz2013,Murmann2015A,Murmann2015B, Dung2017}. Rydberg polaritons, where interactions between photons are mediated by atomic Rydberg states, are also a favorable platform for the comparisons between few- and many-body physics~\cite{Jachymski2016PRL}.

By using a bottom-up approach and increasing one by one the number of ultracold atoms in a quasi-one-dimensional system, it was shown in Ref.~\cite{Wenz2013} that the Fermi sea is formed for $N\geq 4$, where $N$ is the number of atoms. Theoretical works have also detected many-body properties for $N=4$. In studies of thermalization in isolated systems, the Fermi-Dirac distribution was obtained for as few as 4 fermions~\cite{Schnack1996,Flambaum1997,Schnack2000,Izrailev2001}, and in a search for integrable systems composed of particles of unequal masses, chaotic spectrum was found for $N=4$~\cite{Harshman2017}. Another interesting experiment related with the main question of this work is the recent demonstration of Bose-Einstein condensation (BEC) with only 8 photons, which is probably the smallest existing BEC~\cite{WalkerARXIV}. A theoretical construction of few-body models to capture ground-state properties of many-body systems has also been proposed~\cite{Ran2017}.

In this work, we consider primarily a one-dimensional (1D) spin-1/2 model with short-range interactions, where the number $N$ of spin excitations is conserved. It can be mapped into a model of spinless fermions using the Jordan-Wigner transformation~\cite{Lieb1961} or of hard-core bosons using the Holstein-Primakoff transformation~\cite{Holstein1940}. Thus, an excitation in the spin model is equivalent to the presence of a particle in those other models. The corresponding Hamiltonians describe experiments with nuclear magnetic resonance platforms~\cite{Ramanathan2011}, cold atoms~\cite{Bloch2008}, and ion traps~\cite{Jurcevic2014,Richerme2014}. For our spin model to be as generic as possible, we ensure that no local symmetries are present, with the exception of the conservation of the total number of excitations. The case in which half of the chain is filled with excitations is taken as our reference for the many-body limit.

We study static and dynamic properties as $N$ increases from 1, with particular interest in identifying how many excitations are needed for the onset of quantum chaos. In isolated interacting many-body quantum systems, the source of chaos is interparticle interactions. Quantum chaos is a main mechanism for the viability of thermalization, it hinders the transition from a metal to an insulator, and it is related with the fast scrambling of quantum information and the linear growth of entanglement and information entropies.

We verify that, for $N \gsim 4$, the static properties of the spin model with different numbers of excitations become analogous to those of the half-filling case. The shape of the density of states (DOS) becomes close to Gaussian, as typical of many-body quantum systems with two-body couplings~\cite{French1970,Brody1981}, and signatures of quantum chaos, such as the Wigner-Dyson distribution of the spacings between neighboring levels, rigid spectrum~\cite{Guhr1998}, and chaotic eigenstates~\cite{ZelevinskyRep1996,Borgonovi2016} become evident.

Turning to the dynamics, the threshold between few-body and many-body becomes fuzzier. The behavior of the system depends on the time scale and, quite expectedly, on the initial separation between the excitations. For initial states where the excitations are very close to each other, the initial evolution after a quench is similar to what we find for chaotic many-body quantum systems. The Shannon (information) entropy, for instance, grows linearly in time~\cite{Santos2012PRL}. Later, as the excitations spread out, the entropy growth slows down. The behavior of the Shannon entropy becomes logarithmic, similarly to what is seen in disordered many-body systems approaching spatial localization~\cite{Kjall2014,Torres2017}. However, the pre-factors of the logarithms in our clean systems are larger than those in disordered models.

Various tools are available for the analysis of the quench dynamics of 1D systems at the two extreme limits, namely single particle and half filling. The case of $N=1$ is rather trivial, especially in clean models with short-range couplings, while generic properties can often be identified when dealing with chaotic many-body quantum systems~\cite{Flambaum2001b,Torres2017PTR,Torres2018}. We believe that further studies of the region between the two extremes should not only improve our understanding of open problems associated with quantum systems of many interacting particles, such as the quantum-classical correspondence~\cite{Akila2017} and the metal-insulator transition~\cite{Fleishman1980}, but may also reveal new~\cite{Shepelyansky1994} and counterintuitive features~\cite{Santos2003PRB}.

This paper is organized as follows. In Sec.~\ref{sec:model}, we present the Hamiltonian and describe its main features. In the following sections, we analyze standard quantities associated with the eigenvalues and the eigenstates of many-body quantum systems. In particular, in Sec.~\ref{sec:DOS}, we show how the DOS approaches a Gaussian distribution as $N$ increases. Sections~\ref{sec:WD} and~\ref{sec:PR} deal with signatures of quantum chaos. In Sec.~\ref{sec:WD}, we investigate how the correlations between the eigenvalues increase with $N$ leading to level repulsion and rigid spectrum. In Sec.~\ref{sec:PR}, we study the structure of the eigenstates and compare them to those of full random matrices. Section ~\ref{sec:dynamics} concentrates on the dynamics, also employing a quantity of interest in studies of many-body quantum systems. We analyze the growth of the Shannon entropy in time.  Section~\ref{sec:conclusion}  summarizes our results.

\section{System Model} \label{sec:model}

We study 1D spin-1/2 models described by the following Hamiltonian,
\ba
H &=&  J\left[ d_1 S_1^z  +  d_L S_L^z+\epsilon\sum_{i=1}^LS_i^z\right.  \nonumber \\
 &+&   \sum_{i=1}^{L-1}   \left(S^x_iS^x_{i+1}+S^y_iS^y_{i+1}+\Delta S^z_iS^z_{i+1}\right) \nonumber \\
& +& \left. \lambda   \sum_{i=1}^{L-2}   \left(S^x_iS^x_{i+2}+S^y_iS^y_{i+2}+\Delta S^z_iS^z_{i+2}\right)\right].
\label{eq:HXXZ}
\ea
In the above, we set $\hbar=1$. $L$ is the total number of sites, $J$ is a reference energy scale which we set equal to $1$ and $S^{x,y,z}$ represent the spin $1/2$ operators. The Zeeman splittings of all sites are equal and given by $J\epsilon$, except for two impurities placed at the edges of the chain, which have an excess energy $Jd_{1,l}$.  $\Delta$ is the anisotropy parameter and $\lambda$ is the ratio between nearest-neighbors (NN) and next-nearest-neighbors (NNN) couplings. Assuming that the Zeeman splittings were created with a large magnetic field applied to the whole chain and pointing down in the $z$-direction, we can refer to a spin pointing up in $z$ as an excitation.

In the presence of many excitations, the spin model above is a paradigmatic example of many-body quantum systems. When $\lambda=0$ and $\Delta \neq 0$, Hamiltonian (\ref{eq:HXXZ}) represents the XXZ model, which is integrable. We refer to this case as the NN model to distinguish it from the Hamiltonian with $\lambda \neq 0$, which we name NNN model. The latter is no longer integrable. For $\lambda \sim 1$ and many excitations, the NNN model is strongly chaotic, in the sense of showing level statistics equivalent to those of full random matrices. However, signatures of chaos may get concealed if eigenvalues from different symmetry subspaces are mixed~\cite{Santos2009JMP}. To prevent this from happening, we use parameters that avoid most symmetries. The Hamiltonian (\ref{eq:HXXZ}) conserves the total  magnetization in the $z$-direction, ${\cal S}^z = \sum_{l=1}^L S_l^z$, so our studies focus on an individual ${\cal S}^z$ subspace.  The isotropic point $\Delta=1$ is not considered to avoid conservation of total spin. Open boundary conditions are used to break translational invariance. The edge impurities, $d_1 \neq d_L \neq 0$, break parity symmetry and spin reversal symmetry for the case where ${\cal S}^z=0$. Indeed, when  $d_1 \neq d_L \neq 0$, no conservation laws exist in this model, apart from the conservation of energy and total magnetization in the $z$-direction. 

In the following, we denote by $E_{\alpha}$ and $|\psi_{\alpha}\rangle $ the eigenvalues and the corresponding eigenstates of the Hamiltonian. The dimension of a ${\cal S}^z$ subspace with $N$ excitations is given by $D = L!/[N!(L-N)!]$.

The next sections are dedicated to different figures of merit that characterizes the proximity of the NNN model with $N$ excitations to a many-body quantum system. All of them point in the same direction: many-body properties appear for $N\gsim4$. 

\section{Density of States}\label{sec:DOS}

We start our analysis of the spin model by investigating its eigenvalues and look first at the DOS, defined as
\begin{equation}
\rho(E)  \equiv \sum_{\alpha}  \delta (E - E_{\alpha} ) .
\label{eq:DOS}
\end{equation}
The shape of the DOS is not a signature of chaos, but contains information about how many particles are coupled simultaneously. In many-body systems with few-body couplings only, the DOS is known to have a Gaussian form~\cite{French1970,Brody1981}. The spin models described by Eq.~(\ref{eq:HXXZ}) have only two-body couplings. We therefore investigate how the DOS approaches the Gaussian limit, as excitations are added into the system.

\begin{figure}[htb]
\centering
\includegraphics*[scale=0.55]{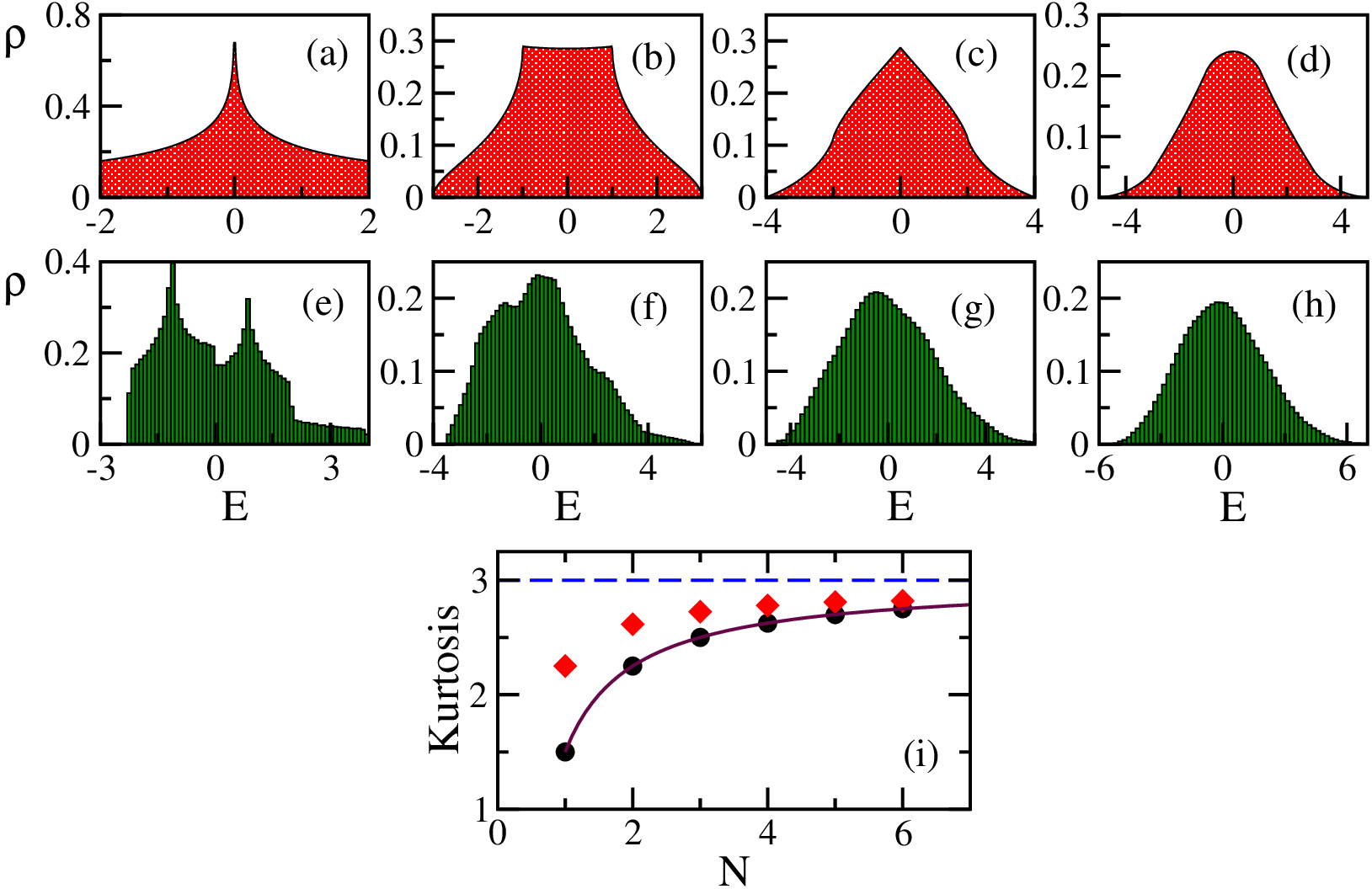}
\caption{Density of states for the XX model (a)-(d) and NNN model (e)-(h). From left to right: $N=2,3,4,5$. For the periodic XX model, we show analytic results in the limit $L\to \infty$. For the NNN model, we take open boundary conditions and system sizes $L=200, 50, 28, 21$ from left to right.  We choose $\Delta=0.48$, $\lambda=1$, $d_1\simeq 0.05$ and $d_L \simeq 0.09$. The energies are rescaled so that for any $L$ and chosen parameters, the middle of the spectrum is at zero. Panel (i) gives the kurtosis of the DOS as a function of $N$ for the XX (circle) and the NNN (diamond) models. The solid line is the fitting curve for the XX model. The kurtosis approaches $3$ as the number of excitations increases.}
\label{Fig:DOS}
\end{figure}

As a warm-up, let us study the case $d_{1,L}=\Delta=\lambda=0$ with closed boundary conditions. This integrable Hamiltonian represents the well-known XX model~\cite{Lieb1961}, for which we are able to compute the DOS exactly. In the continuum limit $L\rightarrow\infty$, as shown in~\ref{app:rhoXX},
\begin{equation}
\rho^{(N)}_{\rm{XX}} (E) =  \frac{1}{2 \pi}\int_{ -\infty }^{\infty}  d\tau e^{iE\tau } {\cal J}_0^N(J\tau ) ,
\label{eq:XXDOS}
\end{equation}
where ${\cal J}_0$ is the Bessel function of the first kind. For $N=1$, this gives trivially 
\be
\rho^{(1)}_{XX}(E)=\frac{1}{\pi}\frac{1}{\sqrt{J^2-E^2}}.
\ee
For $N>1$, the DOS for the XX model is plotted in figures~\ref{Fig:DOS} (a)-(d). From left to right, $N$ grows from 2 to 5. As it can be seen, the peak in the middle of the spectrum becomes progressively smoother and the overall shape of the distribution becomes qualitatively similar to a Gaussian for $N\gsim4$. See also in~\ref{app:extra_data}.1 the DOS for $N=6$ for the integrable XX model, the integrable NN model, and the chaotic NNN model. They all show clear Gaussian shapes.

The approach to the Gaussian shape can be explained in terms of the central limit theorem. As shown in Eq.~(\ref{eq:XX_E}) of~\ref{app:rhoXX}, the eigenvalues of the XX model for $N$ excitations are given by
$E_\alpha=J\sum_{i=1}^N\cos \left( \frac{2\pi k_i}{L} \right)$, where $k_1< k_2<\cdots< k_N$ and $k_i \in\left\{0,\pm1,\pm2,\cdots, \pm(L/2-1),L/2\right\}$. The distribution of the sums of these many cosines is analogous to the distribution of independent random variables, which, according to the central limit theorem, tends to a normal distribution for sufficiently large sample sizes.

A more quantitative way to compare the DOS to a Gaussian distribution is to compute the kurtosis,
\be
K(N)\equiv \frac{\left<(E-\left<E\right>)^4\right>}{\left<(E-\left<E\right>)^2\right>^2},
\ee
where the averages $\langle f(E) \rangle $ are taken on the whole subspace, weighted by the DOS,
\be
\left<f(E)\right>\equiv \int_{-NJ}^{NJ}dE \rho^{(N)}_{XX}(E) f(E).
\ee
For a Gaussian distribution, $K_{G}=3$. In~\ref{app:rhoXX}, we show how to obtain the kurtosis for the XX model,
\ba
K(N)&=&\pi\int_{-\infty}^{\infty}d\tau {\cal J}_0^N(J\tau)\tau^{-5}\left[4NJ\tau(N^2J^2\tau^2-6)\cos (NJ\tau)\right.\nonumber\\
&+&\left.(N^4J^4\tau^4-12 N^2J^2\tau^2+24)\sin (NJ\tau)\right]\label{eq:KXX}\\
&\times&\left(\int_{-\infty}^{\infty}d\tau {\cal J}_0^N(J\tau)\tau^{-3}\left[2NJ\tau\cos (NJ\tau)+(N^2J^2\tau^2-2)\sin (NJ\tau)\right]\right)^{-2}.\nonumber
\ea
The values of $K$ as a function of $N$ are plotted in Fig.~\ref{Fig:DOS} (i). A power-law fit for these points gives
\be
K(N)\sim3\left(1-\frac{1}{2N}\right).
\ee
This tells us that the spectrum smoothly approaches the many-body limit as $N$ is increased, although there is not a threshold at which the limit is reached.

For the NNN model, no exact results are available, so we resort to numerical simulations. From Fig.~\ref{Fig:DOS} (e)-(h), one sees that once again, the shape of the DOS starts looking qualitatively similar to a Gaussian for $N \gsim 4$. In each plot, we have shifted the energies by a constant, such that $\left<E\right>=0$. Notice that we are numerically limited to relatively small sizes, so the system gets denser for $N\gsim 4$. However,  the approach to the Gaussian shape is caused by the increased number of excitations and not by an increasingly dense chain, as made clear by the analysis of the XX model, which is done here in the thermodynamic limit.

The values of the kurtosis for the NNN model with $\lambda=1$ are plotted in Fig.~\ref{Fig:DOS} (i). They are larger than those obtained for the XX model and they also approach the Gaussian limit for $N\gsim 4$.

\section{Onset of Chaos} \label{sec:WD}

The mechanism of quantum chaos in many-body systems is interparticle interaction~\cite{Borgonovi2016}. The results for the DOS in the previous section indicate that $N\gsim4$ can already be considered many. Here and in the next section, we investigate whether this number of excitations is also associated with the onset of quantum chaos. The focus is now entirely on the nonintegrable NNN model.

One of the signatures of quantum chaos is the strong repulsion between neighboring energy levels. In a real symmetric matrix with entries drawn independently from a Gaussian ensemble, that is in a matrix from the Gaussian Orthogonal Ensemble (GOE)~\cite{Guhr1998,MehtaBook}, the spacing $s$ of neighboring unfolded~\cite{Guhr1998} eigenvalues follows the Wigner surmise,
\begin{equation}
P(s)= \frac{\pi s}{2} \exp \left(- \frac{\pi s^2}{4} \right).
\label{eq:goe}
\end{equation}
(For the exact Wigner-Dyson distribution, see Ref.~\cite{MehtaBook}.) An important feature of this distribution is that it vanishes linearly for $s\rightarrow0$, meaning that the probability of two eigenvalues crossing each other is suppressed. This behavior is due to the fact that the eigenvalues are strongly correlated. The spectra of realistic chaotic systems with real and symmetric Hamiltonian matrices also follow the distribution in Eq.~(\ref{eq:goe}).

For sequences of uncorrelated eigenvalues, the distribution of $s$ is Poissonian, $P(s)=\exp(-s)$. The eigenvalues of integrable systems often follow this distribution, because they typically display an extensive set of local conserved quantities, which partition the Hilbert space in many uncorrelated sectors.  We note, however, that in integrable systems with a large number of degenerate levels or with spectra of the ``picket-fence'' type, where the eigenvalues are approximately equally spaced, other distributions are found.

\begin{figure}[htb]
\centering
\includegraphics*[scale=0.55]{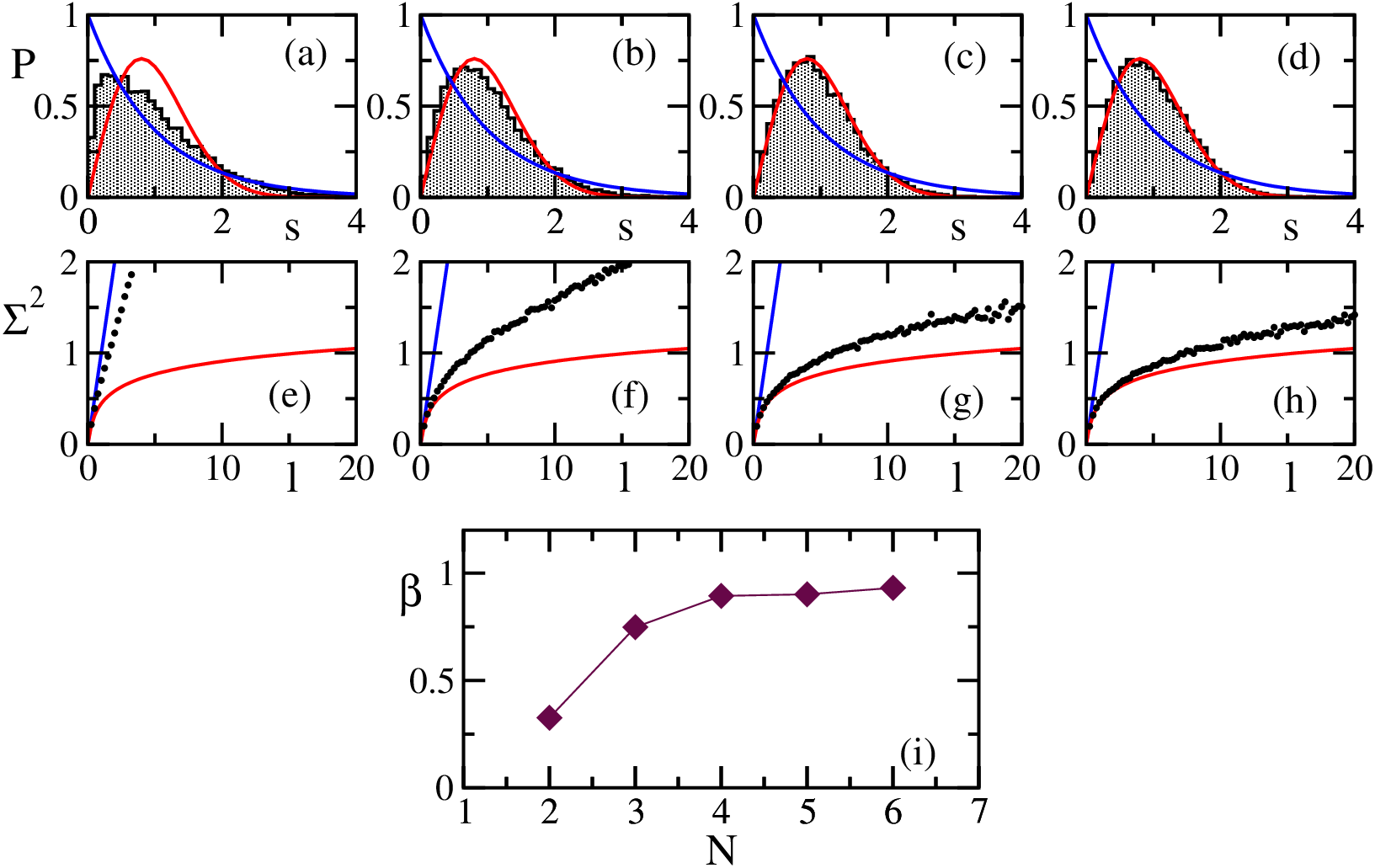}
\caption{Level spacing distribution (a)-(d) and level number variance (e)-(h) for the NNN model. From left to right, the system sizes are $L=200, 50, 28, 21$, respectively, and the numbers of excitations are $N=2,3,4,5$. Parameters: $\Delta=0.48$, $\lambda=1$, $d_1 \simeq 0.05$ and $d_L \simeq 0.09$. Open boundary conditions are taken. In (a)-(d), the grey histogram represents the actual numerical data, which are compared with the Poissonian (blue line) and Wigner-Dyson (red line) distributions. In (e)-(h): numerical data are black dots. They are compared with the result for uncorrelated eigenvalues (blue straight line) and the GOE curve (red logarithmic curve). In (i): the parameter $\beta$ as a function of $N$. It converges to the Wigner-Dyson value $\beta=1$ for $N\gsim4$.}
\label{Fig:Chaos}
\end{figure}

In Fig.~\ref{Fig:Chaos} (a)-(d), we plot the level spacing distribution for the NNN model with $\lambda=1$ for systems with $N=2,3,4,5$ excitations. In each plot, we show for comparison the Poisson (red line) and Wigner-Dyson (blue lines) distributions. For $N=2$,  the distribution is intermediate between Poisson and Wigner-Dyson with a visible dip at small $s$, signaling that some amount of level repulsion is already present in the system. As $N$ is increased, level repulsion becomes enhanced, and at $N=4$ the shape is very close to a Wigner-Dyson distribution.

A way to quantify the transition from Poisson to the Wigner-Dyson distribution is by employing a distribution that interpolates between the two, such as the Brody distribution~\cite{Brody1981},
\begin{equation}
P_\beta(s) \equiv (\beta +1) b s^{\beta} \exp \left( -b s^{\beta +1} \right), \hspace{0.2 cm}
b\equiv \left[\Gamma \left( \frac{\beta + 2}{\beta +1} \right)\right]^{\beta +1},
\label{eq:BrodyParameter}
\end{equation}
where $\Gamma$ is Euler's gamma function. The Poisson distribution corresponds to the case $\beta=0$, while the Wigner-Dyson distribution to $\beta=1$. To evaluate quantitatively the degree of level repulsion, we fit our numerical distribution with $P_\beta(s)$, using $\beta$ as a fitting parameter. The resulting values of $\beta$ are plotted as a function of $N$ in Fig.~\ref{Fig:Chaos}(i). As it can be seen, already at $N=4$ we have $\beta\sim1$, meaning that indeed a system with $4$ excitations can be meaningfully described as chaotic.

Another manifestation of the correlation among energy levels is the rigidity of the spectrum, which can be evaluated with quantities such as the level number variance, which is obtained as follows. We partition the unfolded spectrum in energy intervals of length $\ell$ and compute the number of eigenvalues inside each interval. The variance of the distribution of these numbers is the level number variance $\Sigma^2(\ell)$. For full random matrices from the GOE, we have~\cite{Guhr1998}
\begin{equation}
\Sigma^2 (\ell) = \frac{2}{\pi^2} \left( \ln(2 \pi \ell) + \gamma_e + 1 -\frac{\pi^2}{8} \right) ,
\end{equation}
where $\gamma_e = 0.5772\ldots $ is Euler's constant. In contrast, for systems with an uncorrelated spectrum, one finds $ \Sigma^2 (\ell) =\ell$, while for the harmonic oscillator, one has $ \Sigma^2 (\ell) =0$, due to the complete rigidity of the spectrum. $P(s)$ and $\Sigma^2 (\ell) $ are complementary. The former characterizes the short-range fluctuations of the spectrum and the latter characterizes the long-range fluctuations.

In Fig.~\ref{Fig:Chaos} (e)-(h), we plot the function $\Sigma^2(\ell)$ for the NNN model for $N=2,3,4,5$ excitations. The data (black dots) are compared with the curve for the GOE (red line) and the Poissonian distribution (blue line). For $N=2$, the data are close to the curve for uncorrelated eigenvalues. $N=3$ looks intermediate. For $N\gsim4$, the data at different number of excitations are similar to what we obtain for half-filling (see~\ref{app:extra_data}.2 for $\Sigma^2(\ell)$ for $N=6$ and $N=L/2$), which shows that the rigidity of the spectrum for $N\gsim4$ is equivalent to that found in chaotic many-body quantum systems. It must be noticed that the data follow the GOE curve for small $\ell$ only, for both $N\gsim 4$ and $N=L/2$. This happens because the spectra of realistic chaotic many-body quantum systems are never as rigid as the spectra of full random matrices, where all degrees of freedom interact with each other.

Up to this point, we only considered $\lambda=1$. It is known that in the many-body limit, as $\lambda$ decreases from 1 toward 0, the degree of level repulsion between the eigenvalues decreases until disappearing completely at the integrable point ($\lambda =0$). In ~\ref{app:extra_data}.3, we show that this transition, quantified by the Brody parameter $\beta$ vs $\lambda$, is comparable for $N=5,6$, and $N=L/2$, reinforcing our claim that the properties of the spectrum  for $N\gsim 4$ are already  similar to the case at half filling.

\section{Eigenstates}  
\label{sec:PR}

A complete characterization of a many-body quantum system, and especially determining whether it is chaotic or not, requires also the analysis of its eigenstates. Strongly correlated eigenvalues are directly linked with the onset of nearly ergodic eigenstates. In Sec.~\ref{sec:WD}, we showed that the spectral rigidity for $N\gsim 4$ is equivalent to that for $N=L/2$. We now verify that this is reflected in the structure of the eigenstates as well.

In contrast with the eigenvalues, the study of the eigenstates requires a choice of basis. This choice depends on the physical problem one is interested in. For example, in studies of spatial localization, one employs the site-basis  (also known as computational basis), where on each site the spin either points up or down in the $z$-direction. In the context of quantum chaos, one resorts to the mean-field basis, which corresponds to the eigenstates of the regular (integrable) part of the total Hamiltonian. The term (perturbation) that breaks integrability and brings the system into the chaotic domain, also couples the mean-field basis vectors. The level of complexity of the eigenstates of the Hamiltonian of a quantum many-body system depends on the strength of this term.

In the case of the NNN model, we write the eigenstates  $|\psi_{\alpha} \rangle =  \sum_n C_n^{\alpha} |\phi_n\rangle$ in terms of the eigenstates $|\phi_n\rangle$ of the NN model, whose eigenvalues are denoted by $\varepsilon_n$. The distribution
\begin{equation}
R^{(\alpha)}(E) = \sum_n |C_n^{\alpha}|^2 \delta(E - \varepsilon_n)
\end{equation}
characterizes the spreading of the eigenstate $|\psi_{\alpha} \rangle $ in the mean-field basis. In the many-body limit, the shape of $R^{(\alpha)}(E)$ for states away from the borders of the spectrum broadens from a nearly delta function, when $\lambda \sim 0$, to a Gaussian at strong chaos ($\lambda \sim 1$) \cite{Santos2012PRL}.  In strongly chaotic eigenstates,  the coefficients $C_n^{\alpha} $ are approximately random variables following the Gaussian envelope of the system energy shell. The Gaussian shape of $R(E)$ is a consequence of the Gaussian form of the DOS.

In realistic systems with few-body couplings, only the coefficients $C_n^{\alpha} $ within the energy shell  are non-zero.  This is in contrast with the eigenstates of full random matrices, where all $C_n^{\alpha} $ can be non-zero. The eigenstates of full random matrices are random vectors and therefore fully delocalized in the Hilbert space. One can measure the level of delocalization of the states with quantities such as the participation ratio, defined as
\begin{equation}
PR^{(\alpha)} \equiv \frac{1}{\sum_n |C_n^{\alpha}|^4}.
\end{equation}
For full random matrices from GOE, $PR \simeq D/3$ for any eigenstate. In realistic many-body systems, strongly chaotic states also lead to $PR^{(\alpha)}\propto D$, but the pre-factor is smaller than 1/3.

\begin{figure}[htb]
\centering
\includegraphics*[scale=0.55]{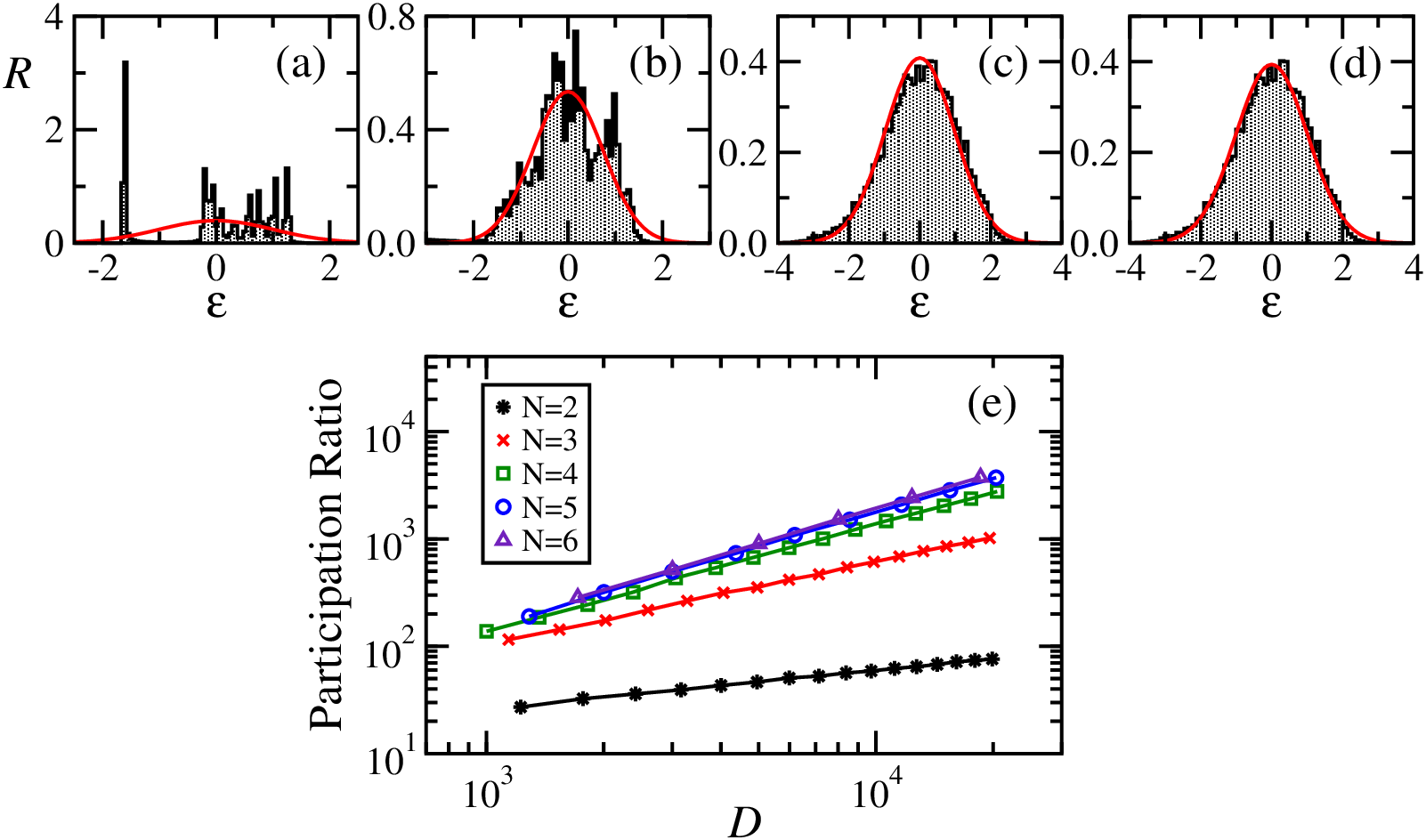}
\caption{Energy distribution $R^{(\alpha)}(E)$ of an eigenstate $\left|\psi_\alpha\right>$ in the middle of the spectrum of the NNN model (a)-(d) and scaling analysis of the participation ratio for eigenstates written in the mean-field-basis (e). Parameters: $\Delta=0.48$, $\lambda=1$, $d_1 \simeq 0.05$ and $d_L \simeq 0.09$. Open boundary conditions are taken. From (a) to (d), the numbers of excitations are $N=2,3,4,5$. The distributions are shifted, such that $\sum_n \left|C^\alpha_n\right|^2 \varepsilon_n =0$, and they are compared with Gaussians (red line) of variance $\sum_n \left|C^\alpha_n\right|^2 \varepsilon_n^2 $. In (e), the data are averaged over 10\% of the eigenstates in the middle of the spectrum. Each curve corresponds to a different $N$ (indicated). For $N \gsim 4$, PR $\propto D$.}
\label{Fig:PR}
\end{figure}

In figs.~\ref{Fig:PR} (a)-(d), we plot the function $R^{(\alpha)}(E)$ for an eigenstate $\left|\psi_\alpha\right>$ near the middle of the spectrum of the NNN model with $\lambda=1$. For this perturbation strength, we know that the eigenstates in the many-body limit are highly delocalized~\cite{Santos2012PRL}. Our goal in Fig.~\ref{Fig:PR} is to analyze how the structure of the eigenstates depends on the number of excitations. For $N=2$, the distribution is sparse, indicating that the eigenstates are far from being fully extended in the energy shell. The level of delocalization increases with $N$. For $N\gsim4$ the distribution already resembles a Gaussian (red curves), indicating the proximity to the chaotic many-body limit. 

To make a more quantitative analysis, in Fig.~\ref{Fig:PR} (e), we show the scaling of $PR$ as a function of the dimension $D$ for different $N$'s. For $N<4$, the scaling is sub-linear, implying that one cannot consider systems with such small number of excitations as fully chaotic. Conversely, for $N\gsim4$, the curves fall closely on top of each other and give $PR\propto D$, indicating that the eigenstates are already very close to the maximal allowed level of spreading over the mean-field basis for the given perturbation strength.

\section{Dynamics}  \label{sec:dynamics}

The characterization of the Hamiltonian developed in the previous sections convinces us that we do not need a large number of excitations to witness properties associated with many-body quantum systems. But are the dynamics of systems with $N\gsim 4$ also comparable to those of $N\sim L/2$? This is a very pertinent question, given the enormous interest in the nonequilibrium dynamics of many-body quantum systems. Information about the dynamics is contained in the eigenvalues and eigenstates, but it depends also on the initial states. In addition, different features of the system may be captured and enhanced at different time scales~\cite{Torres2018,Tavora2016}.

In this section, we analyze the real time evolution of the NNN model with $\lambda=1$. In our simulations, we initialize the system in the following two site-basis states,
\begin{eqnarray}
&&|\Psi^{(N)}_1(0) \rangle  = | \downarrow _1 \cdots \underbrace { \uparrow _j  \uparrow _{j + 1} \cdots  \uparrow _{j + N - 2}  \uparrow _{j + N - 1} }_{N \,{\rm sites}, \,N\,{\rm up-spins} } \cdots \downarrow _L\rangle , \nonumber \\
&&|\Psi^{(N)}_2(0) \rangle  = | \downarrow _1 \cdots \underbrace { \uparrow _j  \downarrow _{j + 1} \cdots \uparrow _{j + 2N - 2} \downarrow _{j + 2N - 1} }_{2N \,{\rm sites},\,N \,{\rm up-spins} } \cdots  \downarrow _L \rangle. \nonumber
\end{eqnarray}
In the half filling limit, $|\Psi^{(L/2)}_1(0) \rangle$ coincides with  the domain wall state and $|\Psi^{(L/2)}_2(0) \rangle $ with the N\'eel state. These states are accessible experimentally~\cite{Trotzky2008,Schreiber2015} and have been extensively investigated in  theoretical studies of quench dynamics in the ${\cal S}^z \!\! =\! 0$ subspace. 

The quantity chosen for the analysis of the dynamics is the Shannon (information) entropy,
\be
S_h(t)\equiv  - \sum \limits_j  |W_{j} (t) | \ln |W_{j} (t) |,
\label{Shan}
\ee
where $W_{j} (t)= |\langle \phi_j | e^{ - iHt} | \Psi (0)\rangle |^2$. This quantity measures how the initial state spreads into other site-basis vectors $|\phi \rangle$ and is related to the entanglement entropy~\cite{Torres2016Entropy}. In many-body quantum systems with highly delocalized initial states, the Shannon entropy is known to increase linearly in time~\cite{Santos2012PRL,Torres2016Entropy}.

\begin{figure}[htb]
\centering
\includegraphics*[scale=0.55]{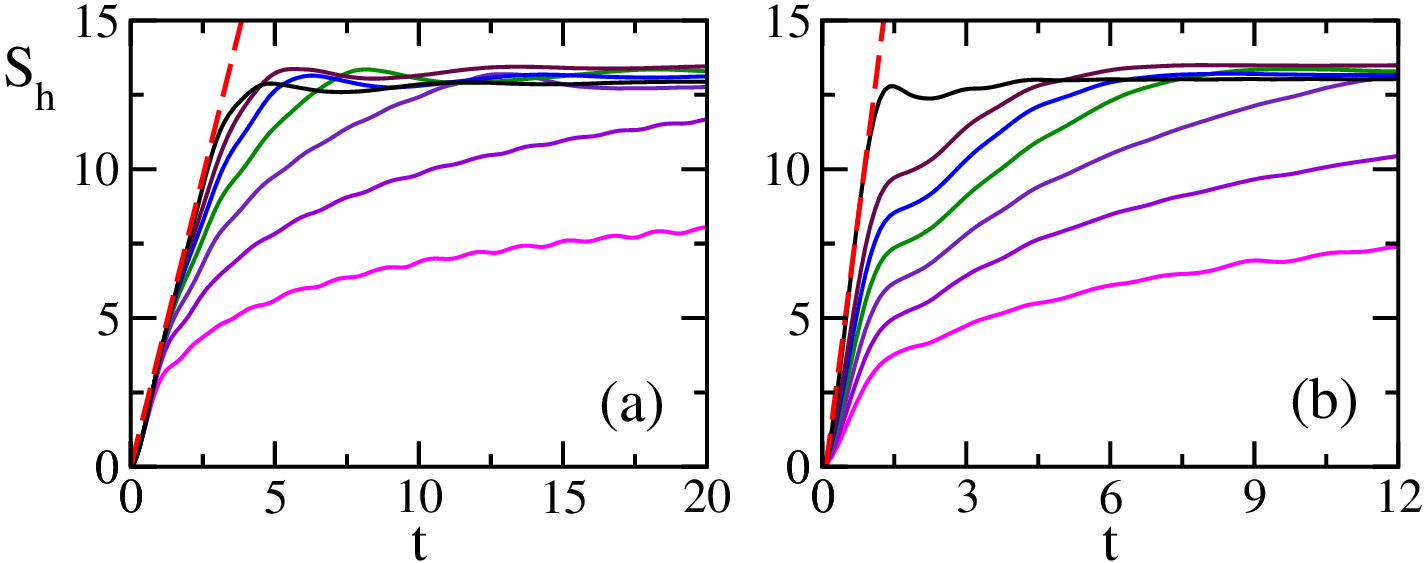}
\caption{Evolution of the Shannon entropy in the site-basis. NNN model with $\Delta=0.5$, $\lambda=1$, periodic boundary conditions, and no impurities. $|\Psi^{(N)}_1(0) \rangle $ (which becomes the domain wall state at half-filling) is considered in (a) and $|\Psi^{(N)}_2(0) \rangle$ (which becomes the N\'eel state at half-filling) is used in (b). From bottom to top, the lines correspond to $(L,N) = (1400,2)$, $(184,3)$, $(72,4)$, $(44,5)$, $(32, 6)$, $(28, 7)$,  and $(22, 11)$ (half filling limit). A linear increase (red dashed line) is shown for comparison with the half filled case.}
\label{Fig:Sh}
\end{figure}

Figure~\ref{Fig:Sh} shows the evolution of $S_h(t)$ for $N=2,3,4,5,6,7$, and also for the half-filling limit $N=L/2$. The system sizes used lead to dimensions $D$ of the Hilbert space that are similar for all $N$'s, so that the saturation point of the dynamics for all curves are of the same order. Data for the initial state $|\Psi^{(N)}_1(0) \rangle$ are shown in Fig.~\ref{Fig:Sh} (a) and those for $|\Psi^{(N)}_2(0) \rangle$ in Fig.~\ref{Fig:Sh} (b). 

As expected, for the state $|\Psi^{(N)}_1(0) \rangle$, $S_h(t)$ follows exactly the half-filling curve up to a time scale dependent on $N$. This happens because the short-time dynamics are restricted to the interface between the up and down spins at the edges of the domain. For the state $|\Psi^{(N)}_2(0) \rangle$, the evolution does not follow the half-filling curve at any time.

In the half-filled case, the two initial states qualitatively evolve in the same way. The Shannon entropy grows linearly, indicating that the site-basis vectors are populated exponentially fast in time, as typical of chaotic systems. For the other $N$'s, the linear growth holds for a certain time interval, but then the evolution slows down and becomes logarithmic. The time interval of the linear behavior increases with $N$, but the crossover between the fast relaxation at short times and the slow dynamics at long times is always visible. This may be interpreted as a crossover from a short-time regime, in which the excitations interact with each other, and a later-time regime, in which the excitations are too diluted to experience strong interactions.  Even though somewhat expected, this change in behavior was not anticipated from the analysis of the eigenvalues and eigenstates developed here. This suggests that with other quantities and more refined analysis, we may be able to distinguish systems with $N=L/2$ from those with $N\neq L/2$ already at the static level. For $N\gsim 4$, the source of these differences should however be associated with the filling of the chain~\cite{Vidmar2017} and not with a small {\em vs.}  large number of excitations. 

We note that a similar change in the dynamic behavior occurs also in many-body models with onsite disorder as they approach localization in space~\cite{Kjall2014,Torres2017}. However, the pre-factor of the logarithmic behavior in this case is smaller than 1 and related with the fractal dimension of the eigenstates~\cite{Torres2017}. The pre-factor in our clean model with few excitations is larger than 2 and it increases with $N$. In fact, at least for $N$ up to $4$, the pre-factor seems to be $\sim N$. 

Overall, the results display interesting features that will be studied in greater detail in a future work. This includes the pre-factor of the logarithmic behavior and how it depends on the number of excitations, the initial states, and the bounds in the energy spectrum.

\section{Conclusion} 
\label{sec:conclusion}

While many tools exist to study systems in the single-particle and in the many-body limit, the crossover between these two regimes is still poorly understood. In this work, we analyzed one of the aspects of this crossover, namely how signatures of quantum chaos emerge as the number of excitations increases. We showed that many-body properties associated with the eigenvalues and eigenstates manifest themselves already for as few as $N\gsim4$. This was done by analyzing different standard indicators of quantum chaos and finding that they all give consistent results. It is interesting that other many-body properties were also found for $N\sim 4$ in experimental~\cite{Wenz2013} and theoretical~\cite{Schnack1996,Flambaum1997,Schnack2000,Izrailev2001, Harshman2017} works.

From the point of view of the dynamics of the system, the behavior depends on the time scale and initial state. If one initially confines all excitations to a small region, the evolution at short times is dominated by the interactions and the behavior is analogous to the many-body case. At long times, the excitations spread out and the effects of the interactions fade away. The analysis of the crossover between the two different temporal regimes is  within reach of existing experiments with cold atoms and ion traps, where the number of particles considered in the dynamics can be manipulated.

The fact that we can detect many-body properties for as few as 4 particles is of course of great relevance for experimental and theoretical studies of many-body quantum systems, as well as to the development of new numerical methods targeting these systems. It implies, for example, that an out-of-equilibrium isolated interacting quantum system with only $N\gsim 4$ should be able to reach thermal equilibrium. It also means that it may be as hard to localize an interacting system with about 4 particles as it is for $N=L/2$. 

It is our hope that this work will motivate further research on how the properties of quantum systems change as the number of particles increases. It would be interesting to extend our studies to non-chaotic systems, such as exactly integrable models, where analytical results could be obtained, and many-body localized systems. For the latter, localization properties may change from few particles to the many-body limit. This analysis may shed light on the influence of finite size effects on the localization transition~\cite{Devakul2015,DeRoeck2016}.

\ack
This work was funded by the American National Science Foundation (NSF) Grant No. DMR-1603418.

\appendix
\section{Analytical derivation of the DOS and kurtosis for the XX model}
\label{app:rhoXX}

Following Refs.~\cite{Alcaraz1987,Alcaraz1988,Santos2016}, we compute the DOS for the XX model with periodic boundary conditions using the coordinate Bethe Ansatz method. 
We recall that the Hamiltonian of the XX model is
\be
H=J\sum_{l=1}^L\left(S^x_lS^x_{l+1}+S^y_lS^y_{l+1}\right).
\ee
The DOS  is defined as
\be
\rho_{XX}(E)\equiv\frac{1}{D}\sum_{\alpha=1}^D\delta\left(E-E_\alpha\right)=\frac{1}{D}\sum_{\alpha=1}^D\int_{-\infty}^{+\infty}\frac{d\tau}{2\pi}e^{i\tau(E-E_\alpha)}.
\ee

\subsection{One excitation}
Since the total magnetization of the system is conserved, let us consider first the case $N=1$, that is,  only one spin pointing up in the positive $z$-direction. 
We write the eigenstates $\left|\psi_\alpha\right>$ in the following form,
\be
\left|\psi_\alpha\right>=\sum_{l=1}^L a_l^{\alpha}\left|\phi_l\right>,
\ee
where $\left|\phi_l\right>$ denotes the states of the site-basis (computational-basis),
\ba
\left|\phi_1\right>&=&\left|\uparrow\downarrow\downarrow\cdots\downarrow\right>\nonumber\\
\left|\phi_2\right>&=&\left|\downarrow\uparrow\downarrow\cdots\downarrow\right>\\
&\cdots&\nonumber
\ea
To find the coefficients $a_l^{\alpha}$, we solve the Schr\"odinger equation
\be
H\left|\psi_\alpha\right>=E_\alpha\left|\psi_\alpha\right>,\label{eq:Schroedinger}
\ee
where $E_\alpha$ is the eigenenergy of state $\left|\psi_\alpha\right>$. The Hamiltonian acts on the states $\left|\phi_l\right>$ as follows,
\ba
H\left|\phi_1\right>&=&\frac{J}{2}\left(\left|\phi_L\right>+\left|\phi_2\right>\right);\nonumber\\
H\left|\phi_l\right>&=&\frac{J}{2}\left(\left|\phi_{l-1}\right>+\left|\phi_{l+1}\right>\right), \quad l\neq1,L;\\
H\left|\phi_L\right>&=&\frac{J}{2}\left(\left|\phi_{L-1}\right>+\left|\phi_1\right>\right).\nonumber
\ea
Substituting these relations into Eq.~(\ref{eq:Schroedinger}) and collecting terms with the same index $l$, we get the equation
\be
E_\alpha a_l^{\alpha}=\frac{J}{2}\left(a_{l-1}^{\alpha}+a_{l+1}^{\alpha}\right).
\label{eq:a}
\ee
We now make the following Ansatz for the coefficients $a_l^{\alpha}$,
\be
a_l^{\alpha}=e^{i\theta l},
\ee
where $\theta$ is a real number yet to be determined. Substituting into Eq.~(\ref{eq:a}), we find that
\be
E_\alpha=J\cos\theta.
\ee
We now invoke the periodic boundary conditions. Since
\be
a_{l+L}=a_{l},
\ee
we have that
\be
\theta=\frac{2\pi k}{L},
\ee
with $k\in\left\{0,\pm1,\pm2,\cdots,\pm\left(\frac{L}{2}-1\right),\frac{L}{2}\right\}$. 

In hands of the eigenvalues, we can obtain the DOS,
\be
\rho^{(1)}_{XX}(E)=\frac{1}{L}\sum_{k}\int_{-\infty}^{+\infty}\frac{d\tau}{2\pi}e^{i\tau \left[ E-J\cos \left( \frac{2\pi k}{L} \right) \right] }.
\ee
For $L\gg1$, we can take the continuum limit, making the substitution
\be
\frac{1}{L}\sum_{k}\rightarrow\int_{-\pi}^{\pi}\frac{dq}{2\pi}.\label{eq:continuum}
\ee
This yields
\ba
\rho^{(1)}_{XX}(E)&=&\int_{-\pi}^{\pi}\frac{dq}{2\pi}\int_{-\infty}^{+\infty}\frac{d\tau}{2\pi}e^{i\tau(E-J\cos q)}\nonumber\\
&=&\int_{-\infty}^{+\infty}\frac{d\tau}{2\pi}e^{iE\tau}{\cal J}_0(J\tau)\\
&=&\frac{1}{\pi\sqrt{J^2-E^2}}.\nonumber
\ea
In the above, ${\cal J}_0$ is the Bessel function of the first kind.

\subsection{More than one excitation}
This treatment can be extended to generic values of $N$. The eigenstates can be written as
\be
\left|\psi_\alpha\right>=\sum_{l_1,l_2,\cdots,l_N=1}^L a_{l_1,l_2,\cdots,l_N}^{\alpha}\left|\phi_{l_1,l_2,\cdots,l_N}\right>,
\ee
and the eigenvalues turn out to be
\be
E_\alpha=J\sum_{i=1}^N\cos \left( \frac{2\pi k_i}{L} \right),\label{eq:XX_E}
\ee
with $k_1<k_2<\cdots<k_N$.
Let us look first at the case $N=2$. The DOS is 
\ba
\rho^{(2)}_{XX}(E)&=&\frac{2}{L(L-1)}\sum_{k_1<k_2}\int_{-\infty}^{+\infty}\frac{d\tau}{2\pi}e^{i\tau \left[ E-J\cos \left( \frac{2\pi k_1}{L} \right) -J\cos \left( \frac{2\pi k_2}{L} \right) \right] }\nonumber\\
&=&\frac{1}{L(L-1)}\sum_{k_1\neq k_2}\int_{-\infty}^{+\infty}\frac{d\tau}{2\pi}e^{i\tau \left[ E-J\cos \left( \frac{2\pi k_1}{L} \right) -J\cos \left( \frac{2\pi k_2}{L}\right) \right] }.\label{eq:g2}
\ea
In order to take the thermodynamic limit for this sum, we need to remove the constraint $k_1\neq k_2$. To do that, we sum and subtract the terms with $k_1=k_2$ from the r.h.s. of Eq.~(\ref{eq:g2}). This gives
\ba
\rho^{(2)}_{XX}(E)&=&\frac{1}{L(L-1)}\sum_{k_1,k_2}\int_{-\infty}^{+\infty}\frac{d\tau}{2\pi}e^{i\tau \left[ E-J\cos \left( \frac{2\pi k_1}{L} \right) -J\cos \left( \frac{2\pi k_2}{L} \right) \right] }\nonumber\\
&-&\frac{1}{L(L-1)}\sum_{k}\int_{-\infty}^{+\infty}\frac{d\tau}{2\pi}e^{i\tau \left[ E-2J\cos \left( \frac{2\pi k}{L}\right) \right]}.
\ea
We can now use Eq.~(\ref{eq:continuum}), which yields
\ba
\rho^{(2)}_{XX}(E)&=&\frac{L}{(L-1)}\int_{-\infty}^{+\infty}\frac{d\tau}{2\pi}\int_{-\pi}^{+\pi}\frac{dq_1}{2\pi}\frac{dq_2}{2\pi}e^{i\tau(E-J\cos q_1-J\cos q_2)}\nonumber\\
&-&\frac{1}{(L-1)}\int_{-\infty}^{+\infty}\frac{d\tau}{2\pi}\int_{-\pi}^{+\pi}\frac{dq}{2\pi}e^{i\tau(E-2J\cos q)}.\label{eq:g2int}
\ea
Finally, we take the limit $L\rightarrow\infty$. In this limit, the second term on the r.h.s. of Eq.~(\ref{eq:g2int}) vanishes, leaving us with the result
\be
\rho^{(2)}_{XX}(E)=\int_{-\infty}^{+\infty}\frac{d\tau}{2\pi}e^{i\tau E}{\cal J}_0^2(J\tau).
\ee

One can repeat this computation for any finite $N$, which gives
\be
\rho^{(N)}_{XX}(E)=\int_{-\infty}^{+\infty}\frac{d\tau}{2\pi}e^{i\tau E}{\cal J}_0^N(J\tau) .
\ee

\subsection{Kurtosis}
To compute the kurtosis
\be
K(N)=\frac{\left<(E-\left<E\right>)^4\right>}{\left<(E-\left<E\right>)^2\right>^2},
\ee
the first step is to compute the mean energy $\left<E\right>$,
\ba
\label{eq:E}
\left<E\right>&=&\int_{-NJ}^{NJ}dEE\int_{-\infty}^{\infty}\frac{d\tau}{2\pi} e^{iE\tau}{\cal J}_0^N(J\tau)\\
&=&2i\int_{-\infty}^{\infty}\frac{d\tau}{2\pi}{\cal J}_0^N(J\tau)\int_0^{NJ}dEE\sin (E\tau ) \nonumber\\
&=&2i\int_{-\infty}^{\infty}\frac{d\tau}{2\pi}{\cal J}_0^N(J\tau)\frac{\sin (NJ\tau )-NJ\tau\cos (NJ\tau)}{\tau^2}.\nonumber
\ea
Since the function ${\cal J}_0$ has even parity, the whole integrand on the r.h.s. of Eq.~(\ref{eq:E}) has odd parity, and therefore its integral from $-\infty$ to $\infty$ vanishes. This means that $\left<E\right>=0$ for any $N$, which is a consequence of the fact that the Hamiltonian is symmetric under the transformation $S_z\rightarrow-S_z$. Analogously, any odd moment $\left<E^{2k+1}\right>$ vanishes as well.

Next, we compute the first two even moments of the energy:
\ba
\left<E^2\right>&=&\int_{-NJ}^{NJ}dEE^2\int_{-\infty}^{\infty}\frac{d\tau}{2\pi} e^{iE\tau}{\cal J}_0^N(J\tau)\\
&=&2\int_{-\infty}^{\infty}\frac{d\tau}{2\pi}{\cal J}_0^N(J\tau)\int_0^{NJ}dEE^2\cos (E\tau ) \nonumber\\
&=&2\int_{-\infty}^{\infty}\frac{d\tau}{2\pi}{\cal J}_0^N(J\tau)\frac{2NJ\tau\cos (NJ\tau ) +(N^2J^2\tau^2-2)\sin (NJ\tau) }{\tau^3},\nonumber
\ea
and
\ba
\left<E^4\right>&=&\int_{-NJ}^{NJ}dEE^4\int_{-\infty}^{\infty}\frac{d\tau}{2\pi} e^{iE\tau}{\cal J}_0^N(J\tau)\\
&=&2\int_{-\infty}^{\infty}\frac{d\tau}{2\pi}{\cal J}_0^N(J\tau)\int_0^{NJ}dEE^4\cos (E\tau ) \nonumber\\
&=&2\int_{-\infty}^{\infty}\frac{d\tau}{2\pi}{\cal J}_0^N(J\tau)\left[\frac{4NJ\tau(N^2J^2\tau^2-6)\cos (NJ\tau) }{\tau^5}\right.\nonumber\\
 &+&\left.\frac{(N^4J^4\tau^4-12N^2J^2\tau^2+24)\sin (NJ\tau) }{\tau^5}\right].\nonumber
\ea
Combining the above two formulae, the expression given in Eq.~(\ref{eq:KXX}) follows.

\section{Numerical results for  $\mathbf{N\geq 6}$}
\label{app:extra_data}

For comparison with the figures shown in the main text, where $N\leq 5$, we provide here some additional figures for cases with more than 5 excitations. We also show how chaos develops as the NNN coupling $\lambda$ is increased from $\lambda=0$.

\subsection{Density of states}

The goal of Sec.~\ref{sec:DOS} was to determine when the Gaussian shape, typical of many-body systems with two-body couplings, first appears as $N$ increases from 1. For $N=5$, the distribution looks already very close to Gaussian, as illustrated in Figs.~\ref{Fig:DOS} (d) and (h), and as suggested also by the value of the kurtosis in Fig.~\ref{Fig:DOS} (i). From $N=5$ on, the result only improves, as indicated by the value of the kurtosis for $N=6$ in Fig.1 (i). In Fig.~\ref{Fig:DOS6} below, we show the DOS for $N=6$ for the integrable XX model, the NN model ($\lambda=0$), and the NNN model with $\lambda=1$. They all show clear Gaussian forms, as expected.

\begin{figure}[htb]
\centering
\includegraphics*[scale=0.55]{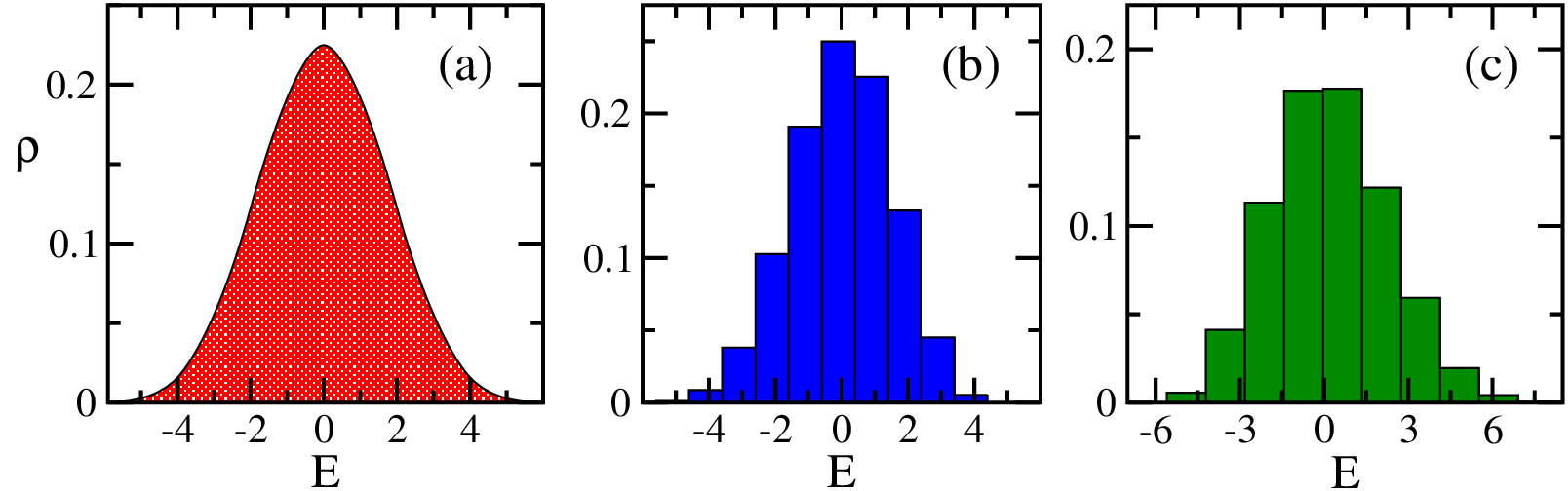}
\caption{Density of states for $N=6$ for the closed XX model (a), open NN model with $\Delta=0.48$ (b) and open NNN model with $\Delta=0.48$ and $\lambda=1$ (c). The DOS for the XX model is computed analytically in the thermodynamic limit using Eq.~(\ref{eq:XXDOS}). For the NN and NNN models, the DOS is obtained numerically for system size $L=18$. The border impurities for the NN and NNN models are $d_1 \simeq 0.05$ and $d_L \simeq 0.09$. In all cases, the Gaussian shape is clearly visible.}
\label{Fig:DOS6}
\end{figure}

\subsection{Level number variance}

\begin{figure}[htb]
\centering
\includegraphics*[scale=0.7]{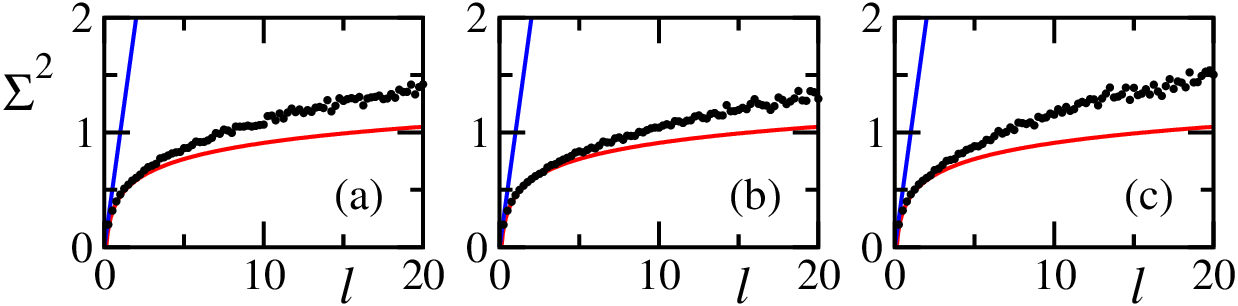}
\caption{Level number variance $\Sigma^2(\ell)$ for $N=5$ and $L=21$ (a), $N=6$ and $L=18$ (b) and half filling, i.e. $N=L/2$ and $L=16$ (c). The numerical data (black dots) are compared with the GOE (red) and Poisson (blue) curves.  Parameters: $\Delta=0.48$, $\lambda=1$, $d_1 \simeq 0.05$ and $d_L \simeq 0.09$.}
\label{Fig:Sigma6}
\end{figure}

In Sec.~\ref{sec:WD}, we showed that the level number variance approaches the GOE curve for $N\gsim 4$. However, one sees that the data for $N=5$ leaves the GOE curve for large $\ell$. This is not caused by the small number of particles and occurs also for $N=6$ and $N=L/2$, as shown in figures~\ref{Fig:Sigma6} (b) and (c). Contrary to the GOE matrices, the Hamiltonian matrices of realistic many-body quantum systems are very sparse and have correlated elements, which results in spectra not as rigid as those of full random matrices. 

The similarities between the results for $N=5$ [Fig.~\ref{Fig:Sigma6} (a)], $N=6$ (b) and $N=L/2$ (c) leave no doubt that for $N\gsim 4$ the system is already comparable to that at half-filling. We notice that, in the many-body limit, the point where $\Sigma^2(\ell)$ escapes the logarithmic curve depends on the dimension of the Hamiltonian matrix and consequent number of available energy levels. The curve for $N=L/2$ in Fig.~\ref{Fig:Sigma6} (c) deviates slightly more from the GOE result than for $N=5$, because for the system sizes considered, $N=L/2$ implies $D=12\,870$ and $N=5$ leads to $D=20\,349$.

\subsection{Crossover to quantum chaos}

\begin{figure}[h]
\centering
\includegraphics*[scale=0.45]{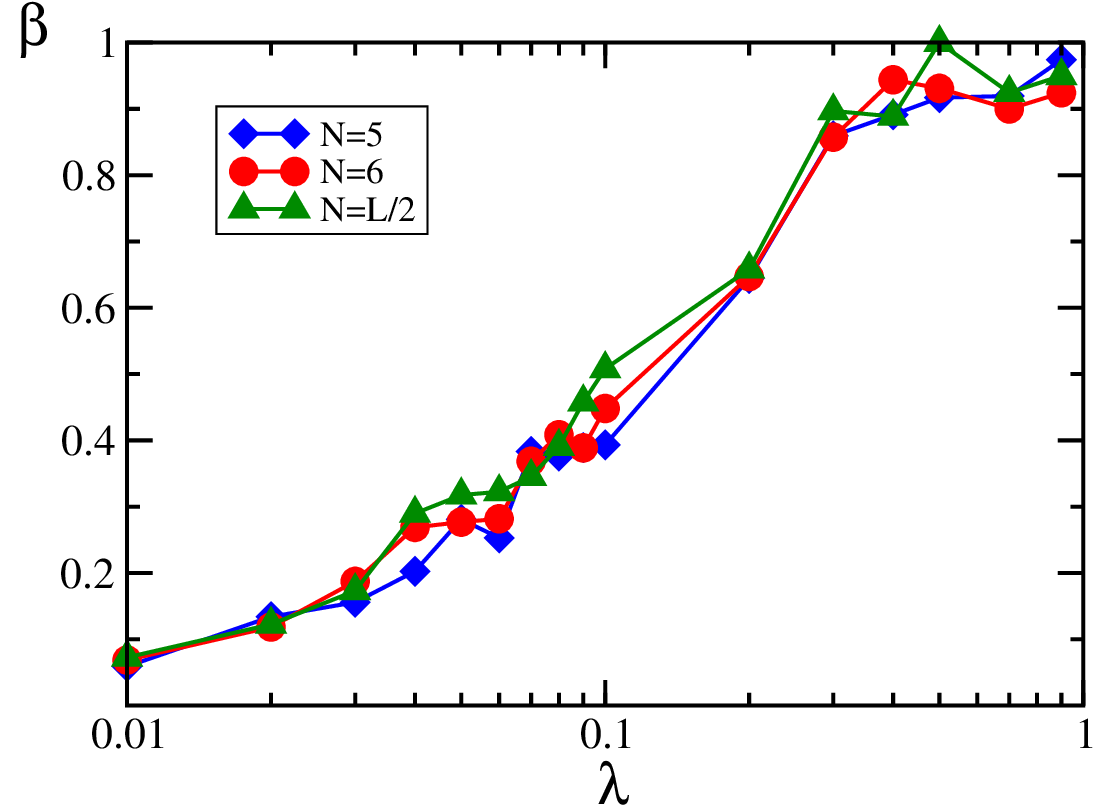}
\caption{Parameter $\beta$ as a function of the NNN coupling $\lambda$ for $N=5$, $N=6$ and half filling, i.e. $N=L/2$. The system sizes are $L=20$, $L=18$ and $L=16$, and the dimensions of the relative Hilbert spaces are $D=15\,504$, $D=18\,564$, and $D=12\,870$, respectively. Parameters: $\Delta=0.48$, $d_1 \simeq 0.05$, and $d_L \simeq 0.09$.}
\label{fig:betavslambda}
\end{figure}

In the many-body limit, it is known that as $\lambda$ in Eq.~(\ref{eq:HXXZ}) decreases below $1$, chaotic signatures associated with the eigenvalues  get progressively blurred and disappear completely in the integrable limit $\lambda=0$ (see {\em e.g.} Ref.~\cite{Torres2014PRE}). A valid question is whether the intermediate degrees of chaoticity, observed for $N\sim L/2$, are equivalent also for small $N\gsim 4$. The answer is yes, as illustrated in Fig.~\ref{fig:betavslambda}. There we show the chaos indicator $\beta$, as defined in Eq.~(\ref{eq:BrodyParameter}), as a function of $\lambda$ for $N=5, 6$ and $N=L/2$. For a fair comparison we consider system sizes $L$ leading to Hamiltonian matrices of similar dimensions $D$, since statistics improves with $D$.

The results for $N=5$, $6$ and $L/2$ look qualitatively similar. 
In all cases, the saturation $\beta\sim1$ is reached around $\lambda\sim0.4$, where the system with $N\gsim 4$ becomes maximally chaotic. 

\vskip 1 cm
\hrule
\vskip 0.5 cm
\bibliographystyle{unsrt}

\begin{thebibliography}{10}

\bibitem{Zinner2016EPJWebConference}
{Zinner, Nikolaj Thomas}.
\newblock Exploring the few- to many-body crossover using cold atoms in one
  dimension.
\newblock {\em EPJ Web of Conferences}, 113:01002, 2016.

\bibitem{Serwane2011}
F.~Serwane, G.~Z{\"u}rn, T.~Lompe, T.~B. Ottenstein, A.~N. Wenz, and S.~Jochim.
\newblock Deterministic preparation of a tunable few-fermion system.
\newblock {\em Science}, 332:336--338, 2011.

\bibitem{Wenz2013}
A.~N. Wenz, G.~Z{\"u}rn, S.~Murmann, I.~Brouzos, T.~Lompe, and S.~Jochim.
\newblock From few to many: Observing the formation of a Fermi sea one atom at
  a time.
\newblock {\em Science}, 342:457--460, 2013.

\bibitem{Murmann2015A}
S.~Murmann, F.~Deuretzbacher, G.~Z\"urn, J.~Bjerlin, S.~M. Reimann, L.~Santos,
  T.~Lompe, and S.~Jochim.
\newblock Antiferromagnetic Heisenberg spin chain of a few cold atoms in a
  one-dimensional trap.
\newblock {\em Phys. Rev. Lett.}, 115:215301, 2015.

\bibitem{Murmann2015B}
S.~Murmann, A.~Bergschneider, V.~M. Klinkhamer, G.~Z\"urn, T.~Lompe, and
  S.~Jochim.
\newblock Two fermions in a double well: Exploring a fundamental building block
  of the {H}ubbard model.
\newblock {\em Phys. Rev. Lett.}, 114:080402, 2015.

\bibitem{Dung2017}
David Dung, Christian Kurtscheid, Tobias Damm, Julian Schmitt, Frank Vewinger,
  Martin Weitz, and Jan Klaers.
\newblock Variable potentials for thermalized light and coupled condensates.
\newblock {\em Nature Photonics}, 11:565, 2017.

\bibitem{Jachymski2016PRL}
K.~Jachymski, P.~Bienias, and H.~P. B\"uchler.
\newblock Three-body interaction of {R}ydberg slow-light polaritons.
\newblock {\em Phys. Rev. Lett.}, 117:053601, 2016.

\bibitem{Schnack1996}
J.~Schnack and H.~Feldmeier.
\newblock Statistical properties of fermionic molecular dynamics.
\newblock {\em Nucl. Phys. A}, 601:181, 1996.

\bibitem{Flambaum1997}
V.~V. Flambaum and F.~M. Izrailev.
\newblock Statistical theory of finite {F}ermi systems based on the structure
  of chaotic eigenstates.
\newblock {\em Phys. Rev. E}, 56:5144, 1997.

\bibitem{Schnack2000}
Hans Feldmeier and J\"urgen Schnack.
\newblock Molecular dynamics for fermions.
\newblock {\em Rev. Mod. Phys.}, 72:655, 2000.

\bibitem{Izrailev2001}
F.~M. Izrailev.
\newblock Quantum-classical correspondence for isolated systems of interacting
  particles: Localization and ergodicity in energy space.
\newblock {\em Phys. Scr.}, T90:95, 2001.

\bibitem{Harshman2017}
N.~L. Harshman, Maxim Olshanii, A.~S. Dehkharghani, A.~G. Volosniev,
  Steven~Glenn Jackson, and N.~T. Zinner.
\newblock Integrable families of hard-core particles with unequal masses in a
  one-dimensional harmonic trap.
\newblock {\em Phys. Rev. X}, 7:041001, 2017.

\bibitem{WalkerARXIV}
Benjamin~T. Walker, Lucas~C. Flatten, Henry~J. Hesten, Florian Mintert, David
  Hunger, A.~A.~P. Trichet, Jason~M. Smith, and Robert~A. Nyman.
\newblock Driven-dissipative, non-equilibrium Bose-Einstein condensation of
  just a few photons.
\newblock arXiv:1711.11087.

\bibitem{Ran2017}
Shi-Ju Ran, Angelo Piga, Cheng Peng, Gang Su, and Maciej Lewenstein.
\newblock Few-body systems capture many-body physics: Tensor network approach.
\newblock {\em Phys. Rev. B}, 96:155120, 2017.

\bibitem{Lieb1961}
Elliott Lieb, Theodore Schultz, and Daniel Mattis.
\newblock Two soluble models of an antiferromagnetic chain.
\newblock {\em Ann. Phys.}, 16:407, 1961.

\bibitem{Holstein1940}
T.~Holstein and H.~Primakoff.
\newblock Field dependence of the intrinsic domain magnetization of a
  ferromagnet.
\newblock {\em Phys. Rev.}, 58:1098, 1940.

\bibitem{Ramanathan2011}
C.~Ramanathan, P.~Cappellaro, L.~Viola, and D.~G. Cory.
\newblock Experimental characterization of coherent magnetization transport in
  a one-dimensional spin system.
\newblock {\em New J. Phys.}, 13:103015, 2011.

\bibitem{Bloch2008}
I.~Bloch, J.~Dalibard, and W.~Zwerger.
\newblock Many-body physics with ultracold gases.
\newblock {\em Rev. Mod. Phys.}, 80:885, 2008.

\bibitem{Jurcevic2014}
P.~Jurcevic, B.~P. Lanyon, P.~Hauke, C.~Hempel, P.~Zoller, R.~Blatt, and C.~F.
  Roos.
\newblock Quasiparticle engineering and entanglement propagation in a quantum
  many-body system.
\newblock {\em Nature}, 511:202, 2014.

\bibitem{Richerme2014}
P.~Richerme, Z.-X. Gong, A.~Lee, C.~Senko, J.~Smith, M.~Foss-Feig,
  S.~Michalakis, A.~V. Gorshkov, and C.~Monroe.
\newblock Non-local propagation of correlations in quantum systems with
  long-range interactions.
\newblock {\em Nature}, 511:198, 2014.

\bibitem{French1970}
J.~B. French and S.~S.~M. Wong.
\newblock Validity of random matrix theories for many-particle systems.
\newblock {\em Phys. Lett. B}, 33:449, 1970.

\bibitem{Brody1981}
T.~A. Brody, J.~Flores, J.~B. French, P.~A. Mello, A.~Pandey, and S.~S.~M.
  Wong.
\newblock Random-matrix physics -- spectrum and strength fluctuations.
\newblock {\em Rev. Mod. Phys}, 53:385, 1981.

\bibitem{Guhr1998}
T.~Guhr, A.~M\"uller-Groeling, and H.~A. Weidenm\"uller.
\newblock Random matrix theories in quantum physics: Common concepts.
\newblock {\em Phys. Rep.}, 299:189, 1998.

\bibitem{ZelevinskyRep1996}
V.~Zelevinsky, B.~A. Brown, N.~Frazier, and M.~Horoi.
\newblock The nuclear shell model as a testing ground for many-body quantum
  chaos.
\newblock {\em Phys. Rep.}, 276:85, 1996.

\bibitem{Borgonovi2016}
F.~Borgonovi, F.~M. Izrailev, L.~F. Santos, and V.~G. Zelevinsky.
\newblock Quantum chaos and thermalization in isolated systems of interacting
  particles.
\newblock {\em Phys. Rep.}, 626:1, 2016.

\bibitem{Santos2012PRL}
L.~F. Santos, F.~Borgonovi, and F.~M. Izrailev.
\newblock Chaos and statistical relaxation in quantum systems of interacting
  particles.
\newblock {\em Phys. Rev. Lett.}, 108:094102, 2012;
L.~F. Santos, F.~Borgonovi, and F.~M. Izrailev.
\newblock Onset of chaos and relaxation in isolated systems of interacting
  spins-1/2: energy shell approach.
\newblock {\em Phys. Rev. E}, 85:036209, 2012.

\bibitem{Kjall2014}
J.~A. Kj\"all, J.~H. Bardarson, and F.~Pollmann.
\newblock Many-body localization in a disordered quantum {I}sing chain.
\newblock {\em Phys. Rev. Lett.}, 113:107204, 2014.

\bibitem{Torres2017}
E.~J. Torres-Herrera and L.~F. Santos.
\newblock Extended nonergodic states in disordered many-body quantum systems.
\newblock {\em Ann. Phys. (Berlin)}, 529:1600284, 2017.

\bibitem{Flambaum2001b}
V.~V. Flambaum and F.~M. Izrailev.
\newblock Entropy production and wave packet dynamics in the {F}ock space of
  closed chaotic many-body systems.
\newblock {\em Phys. Rev. E}, 64:036220, 2001.

\bibitem{Torres2017PTR}
E.~J. Torres-Herrera and L.~F. Santos.
\newblock Dynamical manifestations of quantum chaos: Correlation hole and
  bulge.
\newblock {\em Phil. Trans. R. Soc. A}, 375:20160434, 2017.

\bibitem{Torres2018}
E.~J. Torres-Herrera, Antonio~M. Garc\'{\i}a-Garc\'{\i}a, and Lea~F. Santos.
\newblock Generic dynamical features of quenched interacting quantum systems:
  Survival probability, density imbalance, and out-of-time-ordered correlator.
\newblock {\em Phys. Rev. B}, 97:060303, 2018.

\bibitem{Akila2017}
M.~Akila, D.~Waltner, B.~Gutkin, P.~Braun, and T.~Guhr.
\newblock Semiclassical identification of periodic orbits in a quantum
  many-body system.
\newblock {\em Phys. Rev. Lett.}, 118:164101, 2017.

\bibitem{Fleishman1980}
L.~Fleishman and P.~W. Anderson.
\newblock Interactions and the {A}nderson transition.
\newblock {\em Phys. Rev. B}, 21:2366, 1980.

\bibitem{Shepelyansky1994}
D.~L. Shepelyansky.
\newblock Coherent propagation of two interacting particles in a random
  potential.
\newblock {\em Phys. Rev. Lett.}, 73:2707, 1994.

\bibitem{Santos2003PRB}
L.~F. Santos and M.~I. Dykman.
\newblock Two-particle localization and antiresonance in disordered spin and
  qubit chains.
\newblock {\em Phys. Rev. B}, 68:214410, 2003;
Lea~F Santos and M~I Dykman.
\newblock Quantum interference-induced stability of repulsively bound pairs of
  excitations.
\newblock {\em New J. Phys.}, 14:095019, 2012.

\bibitem{Santos2009JMP}
L.~F. Santos.
\newblock Transport and control in one-dimensional systems.
\newblock {\em J. Math. Phys}, 50:095211, 2009.

\bibitem{MehtaBook}
M.~L. Mehta.
\newblock {\em Random Matrices}.
\newblock Academic Press, Boston, 1991.

\bibitem{Tavora2016}
Marco T\'avora, E.~J. Torres-Herrera, and Lea~F. Santos.
\newblock Inevitable power-law behavior of isolated many-body quantum systems
  and how it anticipates thermalization.
\newblock {\em Phys. Rev. A}, 94:041603, 2016;
Marco T\'avora, E.~J. Torres-Herrera, and Lea~F. Santos.
\newblock Power-law decay exponents: A dynamical criterion for predicting
  thermalization.
\newblock {\em Phys. Rev. A}, 95:013604,  2017.

\bibitem{Trotzky2008}
S.~Trotzky, P.~Cheinet, S.~F\"olling, M.~Feld, U.~Schnorrberger, A.~M. Rey,
  A.~Polkovnikov, E.~A. Demler, M.~D. Lukin, and I.~Bloch.
\newblock Time-resolved observation and control of superexchange interactions
  with ultracold atoms in optical lattices.
\newblock {\em Science}, 319:295, 2008.

\bibitem{Schreiber2015}
Michael Schreiber, Sean~S. Hodgman, Pranjal Bordia, Henrik~P. L\"uschen,
  Mark~H. Fischer, Ronen Vosk, Ehud Altman, Ulrich Schneider, and Immanuel
  Bloch.
\newblock Observation of many-body localization of interacting fermions in a
  quasirandom optical lattice.
\newblock {\em Science}, 349:842, 2015.

\bibitem{Torres2016Entropy}
E.~J. Torres-Herrera, J.~Karp, M.~T\'avora, and Lea~F. Santos.
\newblock Realistic many-body quantum systems vs. full random matrices: Static
  and dynamical properties.
\newblock {\em Entropy}, 18:359, 2016.

\bibitem{Vidmar2017}
Lev Vidmar and Marcos Rigol.
\newblock Entanglement entropy of eigenstates of quantum chaotic {H}amiltonians.
\newblock {\em Phys. Rev. Lett.}, 119:220603,  2017.

\bibitem{Devakul2015}
T.~Devakul and R.~R.~P. Singh.
\newblock Early breakdown of area-law entanglement at the many-body
  delocalization transition.
\newblock {\em Phys. Rev. Lett.}, 115:187201, 2015.

\bibitem{DeRoeck2016}
W.~De~Roeck, F.~Huveneers, M.~M\"uller, and M.~Schiulaz.
\newblock Absence of many-body mobility edges.
\newblock {\em Phys. Rev. B}, 93:014203, 2016.

\bibitem{Alcaraz1987}
F.~C. Alcaraz, M.~N. Barber, M.~T. Batchelor, R.~J. Baxter, and G.~R.~W.
  Quispel.
\newblock Surface exponents of the quantum {XXZ}, Ashkin-Teller and Potts
  models.
\newblock {\em J. Phys. A}, 20:6397, 1987.

\bibitem{Alcaraz1988}
F.~C. Alcaraz, M.~N. Barber, and M.~T. Batchelor.
\newblock Conformal invariance, the XXZ chain and the operator content of
  two-dimensional critical systems.
\newblock {\em Ann. Phys.}, 182:280, 1988.

\bibitem{Santos2016}
L.~F. Santos, M.~T\'avora, and F.~P\'erez-Bernal.
\newblock Excited-state quantum phase transitions in many-body systems with
  infinite-range interaction: Localization, dynamics, and bifurcation.
\newblock {\em Phys. Rev. A}, 94:012113, 2016.

\bibitem{Torres2014PRE}
E.~J. Torres-Herrera and Lea~F. Santos.
\newblock Local quenches with global effects in interacting quantum systems.
\newblock {\em Phys. Rev. E}, 89:062110, 2014.

\end{thebibliography}

\end{document}